\documentclass{aa} 
\usepackage{graphicx}
\usepackage{txfonts}
\begin{document}
\title{BeppoSAX and XMM-Newton spectral study of 4U 1735-44}
\titlerunning{Spectral study of 4U 1735-44}

\subtitle{}
\author{B. M\"uck
					\inst{1},
					S. Piraino\inst{1,2} 
						\and A. Santangelo\inst{1} 
	}

	\institute{$^1$Institute for Astronomy and Astrophysics T\"ubingen, Kepler Center for Astro and Particle Physics, University of T\"ubingen,
		Sand 1, 72076 T\"ubingen, Germany\\
		$^2$INAF-IASF di Palermo, via Ugo La Malfa 153, 90146 Palermo, Italy\\
		\email{benjamin.mueck@uni-tuebingen.de}
	}

\date{}


\abstract
{Low-mass X-ray binary systems consist of a neutron star and a main-sequence companion star. 
	The compact object accretes matter via Roche-lobe overflow, which leads to an accretion disk.
	In addition to a broad-band continuum emission of a thermal component and a Comptonization part, evidence for a broad iron K$_{\alpha}$ line is found in several sources.
Some of them show an asymmetric line profile as well, which could originate from relativistic effects.} 
{To understand the spectral behavior of the system 4U 1735-44, we study the broad-band spectrum and especially the iron line feature between 6.4 and 6.97\,keV. 
The shape of the line allows one to determine the region where the line is produced.
Together with the continuum models, a geometrical model of the source can be proposed. 
Furthermore, the effects of pile-up in the \textit{XMM-Newton} observation are investigated.}
{We analyzed data obtained with the X-ray satellites \textit{BeppoSAX} and \textit{XMM-Newton}.
The \textit{XMM-Newton} data were analyzed, specifically taking into account pile-up effects.
With the help of the data of these two satellites, we performed a detailed spectral study in an energy range from 0.2--24\,keV. }
{During the observations, the source was in the so-called banana state of an atoll source.
Fitting the \textit{BeppoSAX} data, we found line features that we were able to model with a reflection model, whereas the continuum was modeled with a combination of a thermal component and a Comptonization part.
The analysis of the \textit{XMM-Newton} data gave evidence for a broad but not asymmetric iron line.
We found no broadening or asymmetry of the line because of pile-up.}
{}


\maketitle
%

\section{Introduction}

Low-mass X-ray binaries \citep[LMXBs,][]{Hasinger89a} consist of a compact object and a donor companion star with a mass typically not exceeding one solar mass.
In these systems, the compact object is, in most cases, a neutron star, that accretes matter from the donor via Roche-lobe overflow. 
Because of angular momentum conservation, an accretion disk around the neutron star is formed. X-ray emission from LMXBs is complex, and most likely, different components emerge from different regions. 
Generally, the spectrum is well reproduced by one or two thermal components in combination with a Comptonized component at higher energies \citep[e.g.][]{Piraino12a}.
The origin of the different spectral components is still a matter of debate. 
Currently, two main models are widely used to describe the LMXB spectrum: the eastern model \citep{Mitsuda84a, Mitsuda89a} and the western or Birmingham model \citep{Church95a, Church04a}.
\newline
\indent In addition, if the accretion disk is illuminated by radiation from the neutron star, reflection features might appear in the spectrum. 
Indeed, several objects of the class show evidence for broad fluorescence iron K$_\alpha$ lines as well as for other lines, which could originate from reflection from an accretion disk.
Relativistically, broadened fluorescence iron lines were first detected in observations of active galactic nuclei \citep[MCG-6-30-15,][for a review]{Tanaka95, Fabian00a} and in binary systems containing stellar mass black holes \citep[Cyg X-1,][]{Barr85}. Actual relativistic models for lines, such as the well-known disk-line model \citep[][implemented in XSPEC as diskline]{Fabian89a} were introduced to explain the asymmetry of the line observed in Cygnus X-1. More recently, improvements in detection techniques allowed evidence for an asymmetric shape of the iron K$_\alpha$ line also in LMXBs with neutron stars \citep{Piraino1728} to be found. 
To date, a relativistic origin of the fluorescence iron line profile has been discussed in more than ten LMXBs \citep{Bhatt07a,Cackett08a,Cackett09a, DiSalvo09a, Dai10a, Iaria09a, Pandel08a, Piraino12a}.
We have to observe, however, that \cite{DiazTrigo10}, analyzing a large sample of sources observed with \textit{XMM-Newton}, have stressed the importance of the correct instrument pile-up treatment to answer the question whether the iron line is relativistically broadened or not. Very recently, \cite{Piraino12a} and \cite{Egron12a} found that the asymmetry is robust against pile-up effects in the analysis of GX 3+1 and 4U 1705-44, respectively, even when pile-up effects are rigorously taken into account. The superposition of different relativistic effects, such as gravitational redshift and relativistic beaming, shapes the line.
These effects strongly depend on the radius where the line is formed and the inclination under which the system is seen. \\
\indent The presence of such a relativistically broadened line is debated for the LMXB 4U\,1735-44.
The system is located at a distance of 6.5$\pm$1.0\,kpc \citep{GallowayBurst} and consists of a neutron star and a companion star with a mass of 0.53$\pm$0.44\,M$_\odot$ \citep{Casares}. 
Matter is accreted by the compact object via Roche-lobe overflow from the companion, leading to a flux in the 2--10\,keV range of about 4$\times 10^{-9}$\,erg\,cm$^{-2}$\,s$^{-1}$.
The orbital period has been measured to be 4.564$\pm$0.005\,hr by \cite{Corbet89a}. \cite{VanPar88a} first observed irregular type I X-ray bursts. 
Analyzing data of the RXTE satellite, \citet{Wijnands96a,Wijn98a} found a high-frequency quasi-periodic oscillation (QPO) close to 1150\,Hz. 
Subsequently, \citet{Ford173598a} found two simultaneous kHz QPOs with a varying frequency separation of about 300\,Hz.
\cite{Hasinger89a} classified 4U 1735-44 as an atoll source because of its characteristic behavior in a color-color diagram. 
\newline
\indent The broad-band spectrum of the source was discussed by \cite{Seon173597}, who performed a combined analysis of Ginga and ROSAT data.
The spectrum was fitted with a combination of a single blackbody and a Comptonization model. Line features were modeled with Gaussian lines.
Recently, \cite{Lei13a} analyzed RXTE PCA data with the focus on the timing behavior of the source.
The data do not cover the energy range below 2\,keV, and compared with \textit{BeppoSAX} and	\textit{XMM-Newton} the spectral resolution of the PCA instrument is insufficient for investigating the region of the iron line. 
\citet{Cackett09} and \citet{Torr11a} found no evidence for an iron line at all based on Chandra observations at a luminosity range of 2.77--3.37$\times$10$^{37}$\,erg\,s$^{-1}$.  
On the other hand, \cite{DiazTrigo10} found a broad iron K$_{\alpha}$ line, with an equivalent width of 43$\pm$13\,eV and a centroid line energy of 6.74$\pm$0.10\,keV, investigating an 	\textit{XMM-Newton} observation taken at a luminosity of 3.98$\times$10$^{37}$\,erg\,s$^{-1}$. 
\newline
\indent In this paper, we present for the first time a study of the broad-band spectrum of 4U 1735-44 using data of four \textit{BeppoSAX} observations performed in 2000. 
We also report on the \textit{XMM-Newton} observation of a broad iron line in the spectrum of the source, investigating the effects of instrumental pile-up on the line shape. 
We finally suggest a possible source geometry capable of explaining the origin of the observed continuum components and the region where the line formation takes place. 

\section{Observations}

\begin{table}
	\caption{Summary of the analysed observations}             
	\label{BeppoObs}      
	\centering                          
	\begin{tabular}{r c c }        
		\hline\hline                 
		OBSID & Beginning in UT & End in UT 			 \\    
		\hline                        
		\multicolumn{3}{c}{BeppoSax} \\
		\hline                        
		2083600200 & 03/20/2000 17:03:58 &  03/21/2000 14:21:58  \\      
	2122400300 & 08/26/2000 02:59:56 &  08/27/2000 04:24:45  \\
 2122400400 & 09/25/2000 06:31:21 &  09/25/2000 16:04:16  \\
 2122400410 & 09/27/2000 12:41:39 &  09/27/2000 23:39:30  \\
		\hline                                   
		\multicolumn{3}{c}{XMM-Newton}						\\
		\hline                        
		0090230201	& 09/03/2001 03:02:28 & 09/03/2001 09:05:18  \\
		\hline
	\end{tabular}
\end{table}

We analyzed data obtained with the X-ray observatories \textit{BeppoSAX} and \textit{XMM-Newton}.
The observations are summarized in Table \ref{BeppoObs}. 
To extract the data products, we followed the corresponding user handbooks of \textit{XMM-Newton}\footnote{http://xmm.esac.esa.int/} and \textit{BeppoSAX}\footnote{http://www.asdc.asi.it/bepposax/software/index.html}. 
The sax tools that were included in the HEASOFT v.5.3\footnote{http://heasarc.nasa.gov/lheasoft/} were used to extract the \textit{BeppoSAX} data.  
The \textit{XMM-Newton} data products were extracted with the help of the Science Analysis Software (SAS) version 12.0\footnote{http://xmm.esac.esa.int/sas/}. \\
\indent \textit{BeppoSAX} consists of four narrow field instruments, the Low Energy Concentrator Spectrometer \citep[LECS,][]{Parmar97}, the Medium Energy Concentrator Spectrometer \citep[MECS,][]{Boella95a}, the High Pressure Gas Scintillation Proportional Counter \citep[HPGSPC,][]{Giarrusso95a} and the Phoswich Detector System \citep[PDS,][]{Frontera95a}. 
They allow broad-band X-ray observations in an energy range from 0.2--200\,keV, with an effective area of 150\,cm$^2$ at 6\,keV (MECS). \\
\indent Unfortunately, we were unable to use the PDS data in our analysis because of systematic effects above 20\,keV.
These are due to the location of 4U\,1735-44 close to the galactic plane.
During the off-pointing of the PDS collimator, the galactic plane fell into the field of view, resulting in an increased number of background-contaminating sources. \\
The LECS and MECS spectra and lightcurves were used and spectra for the two instruments were extracted from an 8\arcmin and 4\arcmin region around the source center. 
The MECS units 2 and 3 were combined and the MECS unit 1 was not used because it developed problems with the voltage supply in 1997. \\
The four \textit{BeppoSAX} observations have a total exposure time of $\sim$40\,ks for the LECS, $\sim$120\,ks for the MECS and $\sim$82\,ks for the HPGSPC.  
The LECS and MECS background spectra were extracted from event lists of empty-field observations that are provided by the instrument teams.
Since in three of the four observations the collimator of the HPGSPC was not rocking anymore, we summed spectra obtained during Earth occultation to produce a background spectrum, as suggested by the hardware team. This was compared with the off-minus spectrum, the standard background, of the first observation and matched very well. The summed background spectrum provided much better statistics and was therefore used for background subtraction in all four observations.
\indent The observation performed with the EPIC-PN instrument of \textit{XMM-Newton} \citep{Strueder01a} had an exposure time of 5.3\,ks. 
The EPIC-PN CCD detector is sensitive in the range of 0.1--15\,keV and is placed in the focus of an X-ray telescope that offers an effective area of 1500\,cm$^2$ at 6\,keV.
To cope with the expected high count rate of the source, the EPIC-PN camera was operated in timing mode, which offers a time resolution close to 30\,$\mu$s.
In timing mode, ten pixels are collapsed to one macropixel, with the spatial resolution limited to one dimension. 
Despite the relatively high time-resolution of the EPIC-PN camera, pile-up effects had to be properly treated given the high flux of the source.
The inner region of the source should be removed from the extraction, if an observation is affected by pile-up, following the suggestion of the hardware team,.
\begin{figure}
	\centering
	\includegraphics[width=\columnwidth]{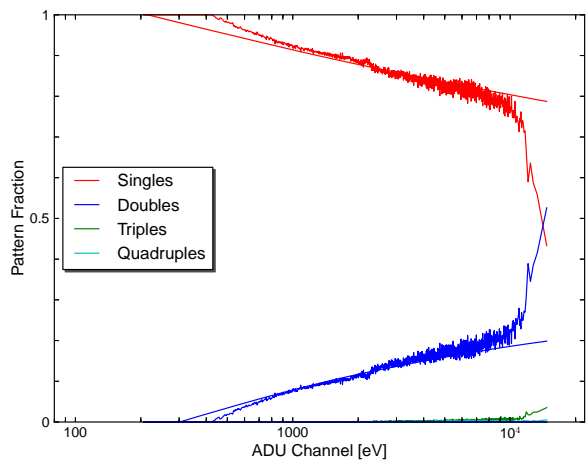}
	\caption{Epatplot of region 1 (All columns). Shown are the fractions of the different pattern types (singles, doubles, triples, and quadruples. The observed pattern fractions are compared with the expected ones (smooth lines). A deviation is an indication of pile-up. Above 7\,keV a clear excess of double events can be seen.}
	\label{Pat15col}
\end{figure}
\begin{figure}
	\centering
	\includegraphics[width=\columnwidth]{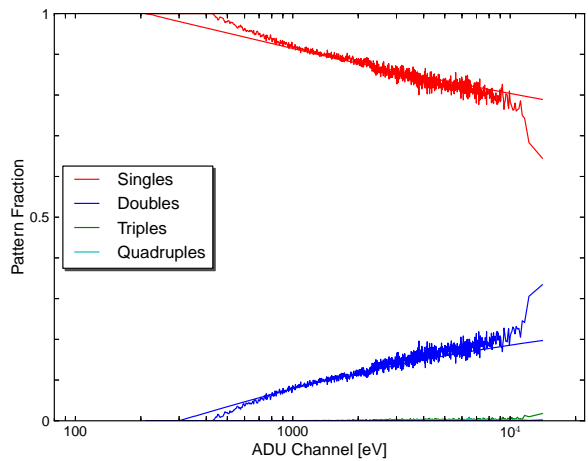}
	\caption{Epatplot of region 3 (without the innermost three columns). Above 10\,keV, an excess of double events is still visible.}
	\label{Pat15_3col}
\end{figure}

To investigate the presence of pile-up we show in Figure \ref{Pat15col} and \ref{Pat15_3col} the so-called \textit{epatplot} results obtained with the SAS analysis software. The \textit{epatplot} shows the fractions of the different pattern types compared with those expected from a pile-up free observation.
As can be seen in Figure \ref{Pat15col}, a clear deviation from a pile-up free observation is observed for data extracted within a box with a width of 15 columns (region 1) centered on the source position (column 38). 
This is indeed expected for a count rate of $\sim$1146\,counts per second.
In a second step, we excluded the three inner columns (37, 38, and 39) of the box in the extraction. 
The epatplot shown in Figure \ref{Pat15_3col} still showed small deviations from the expected pattern distribution above 10\,keV. 
The resulting count rate of $\sim$470\,counts per second is in the border area in which pile-up effects occur.
To compare the effects of pile-up, we also extracted a spectrum without the inner five columns.
For this extraction region, there is no evidence for pile-up in the epatplot.
Finally, we extracted the spectra from four different extraction regions, region 1 with all columns used, region 2 without the innermost column, region 3 without three columns, and region 4 without five columns.
To extract the spectra, we used single and double events in an energy range of 0.2--12\,keV.
The response and the ancillary response file were produced with rmfgen and arfgen, respectively, following the instructions for piled-up observations in the SAS user handbook. \\
The other two cameras onboard \textit{XMM-Newton} \citep[the MOS cameras,][]{Turner01a} were both operated in partial window mode, which can handle a source flux of two counts per second.
We checked the data for pile-up and found that the observation is piled-up to an extremely high degree, therefore we were unable to use the MOS data in our analysis. 
\section{Data analysis and observational results}
The spectral fittings were all performed using the spectral analysis package XSPEC v.12.7.0 \citep{Arnaud96a}.
\subsection{BeppoSAX}
\label{BeppoSec}
\begin{figure}
	\centering
	\includegraphics[width=\columnwidth]{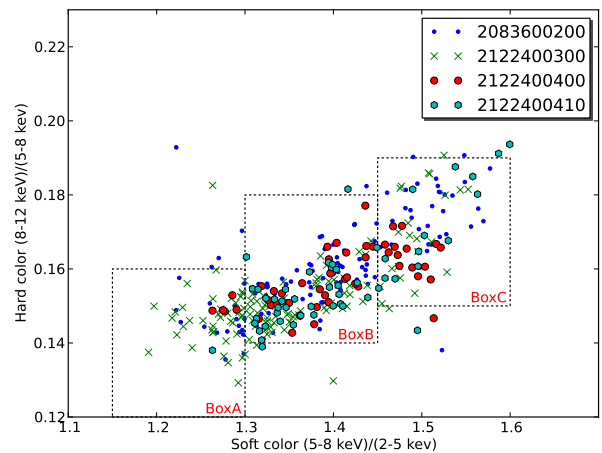}
	\caption{Color-color diagram for the \textit{BeppoSAX} observation. One point represents 400\,seconds of observation. The boxes are discussed at the end of section \ref{BeppoSec}.}
	\label{BeppoCCD}
\end{figure}

\begin{figure}
	\centering
	\includegraphics[width=\columnwidth]{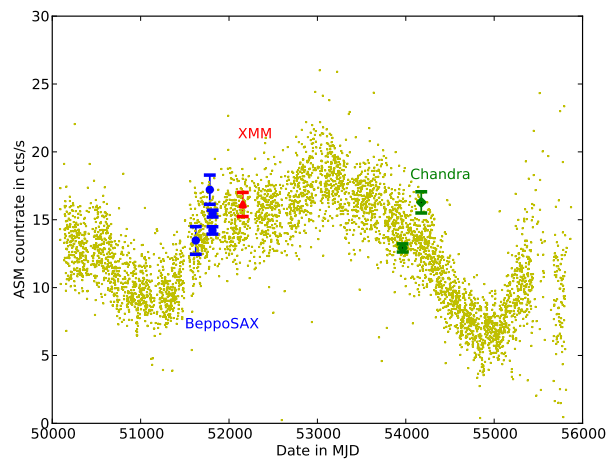}
	\caption{Whole RXTE ASM light curve. The \textit{BeppoSAX} observations are marked with blue points and the \textit{XMM-Newton} with red triangles. The Chandra observations analyzed by \cite{Cackett09} are marked with green diamonds to compare the luminosities.}
	\label{ASM}
\end{figure}

\begin{figure}
	\centering
	\includegraphics[width=\columnwidth]{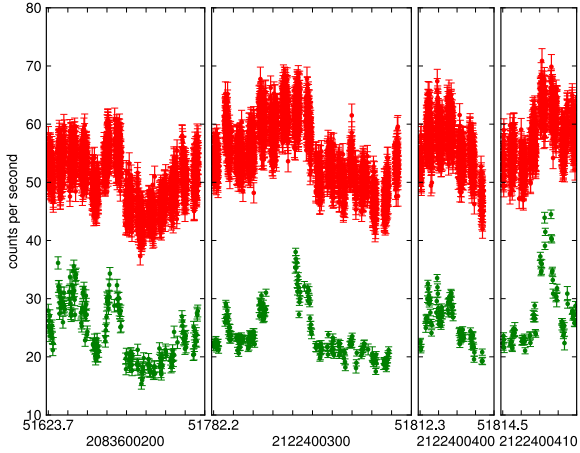}
	\caption{Light curve of the MECS detector with 16\,-second time binning (red) and the HPGSPC light curve with a 196\,-second time binning (green).}
	\label{MecsHPLc}
\end{figure}

To study the spectral variation as a function of the source state, we produced a color-color diagram (Figure \ref{BeppoCCD}) of all observations, plotting the hard color (Flux(7--10\,keV)/Flux(3--7\,keV)) against the soft color (Flux(3--7\,keV)/Flux(1--3\,keV)). During all observations, the source was in the so-called banana state of the atoll sources.
To check time variability on timescales shorter than those of the color-color diagram, light curves of the MECS and HPGSPC data were produced.
As can be seen in Figure \ref{MecsHPLc}, no bursts or dips were found.
Although slight changes in the count rate are observed, transitions between different states cannot be unequivocably identified.\\ 
\newline
\indent To study the intensity of the source on long timescales, the overall light curve of the All Sky Monitor \citep[ASM, ][]{Levine_ASM96} onboard RXTE is plotted in Figure \ref{ASM}. 
The ASM data indicate that the source was in a similar intensity state during the BeppoSAX and the XMM observations, in 2000 and 2001 respectively.
Furthermore, the Chandra observations, performed in 2006 and 2007, were also taken at a similar ASM intensity and were found to be in the banana state by \cite{Cackett09}. We are therefore confident that all observations were performed when the source was in the banana branch. 

\begin{figure}
	\centering
	\includegraphics[width=\columnwidth]{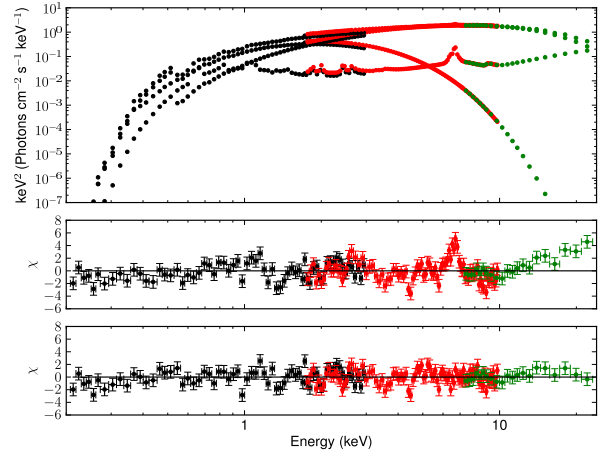}
	\caption{\textit{BeppoSAX} observation 2083600200 with three instruments: LECS (black), MECS (red), and HPGSPC (green). In the upper panel the unfolded spectra multiplied by E$^2$ of the best-fit model (\textsc{phabs*(diskbb + nthcomp + reflionx)}) are shown. In the middle panel the residuals above the continuum are plotted, and in the lower panel the residuals after adding the \textsc{reflionx} model to the continuum.}
	\label{BeppoObs2}
\end{figure}

\indent 
For the LECS spectrum, we used the energy bins from 0.2--3.0\,keV, for the MECS  1.7--10.0\,keV, and for the HPGSPC from 7.0--24.0\,keV.
Above 24\,keV, the HPGSPC data were dominated by the background, so we did not use them.
To cope with the uncertainties of the calibration and  background subtraction, a systematic error of 2\% was applied to the HPGSPC data. \\
\indent The broad-band continuum was modeled with a photoelectric absorbed (\textsc{phabs}) combination of a multicolor disk blackbody \citep[\textsc{diskbb},][]{Mitsuda84a} and a thermally Comptonized component \citep[\textsc{Nthcomp},][]{Zdziarski_nthcomp99a, Zycki_nthcomp99a}. 
The shape of the disk blackbody is determined by the temperature at the inner disk radius T$_{in}$. 
Given the source distance, it is possible to calculate the inner disk radius from the normalization defined as k = (R$_{in}$[km]\,/\,D[10\,kpc])$^2 cos \theta$.
The Comptonization component depends on the power-law photon index $\Gamma$, the electron temperature kT$_e$, the seed photon temperature kT$_{bb}$, and the normalization.  
Furthermore, the \textsc{Nthcomp} model allows one to choose between blackbody and disk blackbody seed photons.  \\
\indent The fit of observation 1 resulted in a reduced $\chi^2$ of 2.39 ($\chi^2$/DoF = 409/171) and some line-like features appeared in the residuals together with an excess above 10 keV. 
The most prominent feature appeared at energies between 5 and 7\,keV, as clearly seen in the middle panel of Figure \ref{BeppoObs2}.
Since reflection from the accretion disk is observed in LMXBs, we used the \textsc{reflionx} model \citep{Ross05a}, which models reflection by an optically thick atmosphere, which approximates the surface of an accretion disk well. In fact, the reflection model has recently been used in the analysis of other LMXBs \citep[see][]{Reis09, Egron11a}. The spectrum of the reflection depends on the abundance of iron relative to the solar value (\textsc{Fe/solar}). The abundances of the other elements are set to their solar values \citep{Morrison83a}. 
The model also depends on the photon index $\Gamma$ of the illuminating power-law spectrum, which was fixed in our fits to the $\Gamma$ of the \textsc{Nthcomp} because we expect that the illuminating photons originate from regions close to the neutron star. Another free-fit parameter is the ionization, defined as $\xi=4\pi F/n$, where F is the total illuminating flux and n the hydrogen number density.
The best-fit results for the four \textit{BeppoSAX} observations are summarized in Table \ref{BeppoFits} and the best-fit residuals as well as the unfolded spectrum are shown in Figure \ref{BeppoObs2}. 
\begin{table}
	\renewcommand{\arraystretch}{1.15}
	\caption{Best-fit results for the four \textit{BeppoSAX} observations}             
	\label{BeppoFits}      
	\centering                          
	\begin{tabular}{l c c c c}        
		\hline\hline                 
		Parameter 								& Obs 1 										& Obs 2 										& Obs 3 										& Obs 4			 \\    
		\hline                        
		N$_H$[10$^{22}\,cm^{-2}$]	& 0.33$^{+0.01}_{-0.02}$		&	0.33$^{+0.02}_{-0.02}$		&	0.30$^{+0.01}_{-0.01}$		&	0.37$^{+0.03}_{-0.03}$		\\
		\hline
		 kT$_{dbb}$[keV]		& 0.66$^{+0.07}_{-0.16}$		& 0.83$^{+0.03}_{-0.03}$		& 0.76$^{+0.07}_{-0.06}$		& 0.80$^{+0.04}_{-0.03}$		\\
			k$_{dbb}$						&	432$^{+430}_{-115}$				&	251$^{+34}_{-28}$					&	288$^{+90}_{-70}$					&	291$^{+46}_{-49}$					\\
 		\hline
		 $\Gamma$				& 1.62$_{fixed}$						&	2.19$_{fixed}$						&	1.78$_{fixed}$						&	2.31$_{fixed}$						\\
		kT$_e$[keV]		&	2.60$^{+0.04}_{-0.04}$			&	3.20$^{+0.14}_{-0.14}$		&	2.80$^{+0.06}_{-0.05}$		&	3.42$^{+0.15}_{-0.15}$		\\
		kT$_{bb}$[keV]&	0.81$^{+0.14}_{-0.31}$			&	1.34$^{+0.04}_{-0.04}$		&	1.05$^{+0.10}_{-0.04}$		&	1.38$^{+0.04}_{-0.04}$		\\
		k$_{comp}$[10$^{-2}$]& 13.6$^{+6.0}_{-3.6}$				&	7.3$^{+0.6}_{-0.5}$				&	10.2$^{+2.0}_{-2.0}$			&	7.6$^{+0.6}_{-0.6}$	\\
		\hline
		Fe/solar			& 1.69$^{+0.51}_{-0.43}$		& 0.94$^{+0.64}_{-0.21}$	&	3.23$^{+4.78}_{-2.10}$			& 1.40$^{+1.46}_{-0.57}$		\\
		$\xi$						&	298$^{+132}_{-43}$				& 550$^{+139}_{-207}$			&	1996$^{+523}_{-730}$					&	559$^{+192}_{-198}$				\\
		k$_{ref}$[10$^{-6}$]	&	13.5$^{5.6}_{-5.5}$				& 5.33$^{+4.61}_{-2.07}$	& 0.94$^{+0.87}_{-0.33}$			& 7.07$^{+3.00}_{-3.25}$		\\
		\hline
		$\chi^2$												&	212												& 194											& 197													& 193								\\
			 D.o.f													& 168											& 167											& 167													& 171												\\
		red. $\chi^2$										& 1.262											&	1.159										&	1.179												&	1.131											\\
		\hline\hline	
	\end{tabular}
\end{table} 

For all four observations, statistically meaningful fit results were obtained.
All line-like features were properly modeled and the residuals showed no need for an additional model component. 
We also tried to model the line-like features with Gaussian lines, obtaining higher values of $\chi^2$ and therefore concluding that the \textsc{reflionx} model describes the source spectrum better.
We also searched for spectral variations along the path of the color-color diagram.
Three spectra were obtained by summing the individual spectra of the three boxes defined in Figure \ref{BeppoCCD}. 
The best-fit results slightly changed, but are still within the error range. Hence, no significant spectral variation was detected as a function of the source position in the color-color diagram.

\subsection{XMM-Newton}

\begin{figure}
	\centering
\includegraphics[width=\columnwidth]{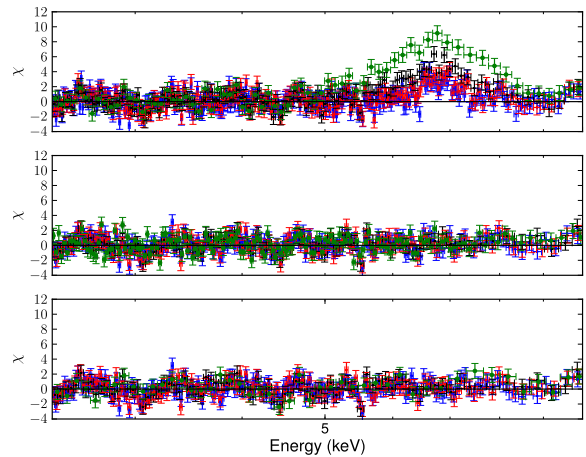} 
\caption{\textit{Upper panel:} The residuals of all four extraction regions obtained by just using the continuum model. Shown here are region 1 (all columns, blue), region 2 (without one column, red), region 3 (without three columns, black) and region 4 (without five columns, green). \textit{Middle panel:} A Gaussian is added to the continuum model. \textit{Lower panel:} The \textsc{diskline} model is used instead of a Gaussian.}
	\label{XMMunf}
\end{figure}

For the nominal energy range of 0.7--10\,keV of the EPIC-PN data, the fit showed a broad feature at $\sim$1\,keV. This feature was found in each of the sources analyzed by \cite{DiazTrigo10}.
Additionally, features are present in the range of Si and Au absorption, which were adressed in an EPIC-PN calibration note\footnote{http://xmm2.esac.esa.int/docs/documents/CAL-TN-0083.pdf}.
To avoid effects on the fit parameters, we decided to remove the energy bins below 2.4\,keV as was done before by e.g. \cite{Piraino12a}.
In addition, energy bins above 10\,keV were not used, as is standard in EPIC-PN observations. 
\newline
\indent The best fit of the continuum spectral component was achieved by combining an absorbed multicolor disk blackbody \textsc{diskbb} and a single blackbody \textsc{bbody}.
Replacing the single blackbody with a Comptonization model, such as \textsc{nthcomp} or \textsc{compTT} \citep{Titarchuk94a}, provided an unsatisfactory fit.
This is not surprising given the limited energy range of the EPIC-PN data, which leads to problems in constraining the parameters.
After modeling of the continuum, line-like residuals are apparent in the energy range of 5--7\,keV, as shown in the upper panel of Figure \ref{XMMunf} for all four chosen extraction regions.
Unfortunately, we could not use the \textsc{reflionx} model to describe the line features given the limited energy range of the \textit{XMM-Newton} data.
This is why we used a simple Gaussian model. The resulting residuals are shown in the middle panel of Figure \ref{XMMunf}.
This model gave acceptable fits with a reduced $\chi^2$ close to 1 for all four regions.
Similar statistically acceptable results were also obtained using the model for relativistically broadened iron lines \textsc{diskline} \citep{Fabian89a} (see residuals in the lower panel of Figure \ref{XMMunf}).
Since we obtained similar results with both models, we cannot distinguish a relativistic line profile with respect to a simpler Gaussian broadening. 
It is worth noting that the radius at which the line is produced is larger than 15 Schwarzschild radii for the \textsc{diskline} model.  
The photoelectric absorption was fixed to 0.30$\times$10$^{22}\,cm^{-2}$, the value was obtained with the help of the \textit{BeppoSAX} data since we did not use energy bins below 2.4\,keV, which leads to problems in defining the absorption. 
The best-fit results for the four different extraction regions are summarized in Table \ref{XMM_4regions}. 
\begin{table}
	\renewcommand{\arraystretch}{1.15}
	\caption{Best-fit results for the four \textit{XMM-Newton} extraction regions. The spectra were extracted from region 1 (all columns), region 2 (without one column), region 3 (without three columns) and region 4 (without five columns). The absorption was frozen to 0.3$\times10^{22}\,cm^{-2}$.}             
	\label{XMM_4regions}      
	\centering                          
	\begin{tabular}{l c c c c}        
		\hline\hline                 
		Parameter 								& Reg 1 										& Reg 2 										& Reg 3 										& Reg 4			 \\    
		\hline                        
		 kT$_{dbb}$[keV]		& 1.00$^{+0.03}_{-0.03}$		& 1.00$^{+0.03}_{-0.02}$			& 0.98$^{+0.04}_{-0.04}$		& 0.91$^{+0.06}_{-0.05}$			\\
		 k$_{dbb}$						&	147$^{+14}_{-16}$					&	157$^{+17}_{-16}$					&	172$^{+27}_{-25}$					&	211$^{+50}_{-45}$						\\
 		\hline                                                                                                                                
		kT$_{bb}$[keV]		&	2.04$^{+0.03}_{-0.02}$		&	2.06$^{+0.03}_{-0.03}$			&	1.99$^{+0.03}_{-0.04}$		&	1.85$^{+0.04}_{-0.03}$			\\
			k$_{bb}[10^{-2}]$							&	5.8$^{+0.1}_{-0.1}$	&	6.4$^{+0.1}_{-0.1}$		&	6.3$^{+0.1}_{-0.1}$	&	5.9$^{+0.1}_{-0.2}$		\\
		\hline                                                                                                                                
		E$_{gau}$[keV]		&	6.79$^{+0.09}_{-0.09}$		&	6.82$^{+0.09}_{-0.09}$			& 6.77$^{+0.09}_{-0.10}$ 		&	6.70$^{+0.15}_{-0.13}$ 			\\
		$\sigma_{gau}$[keV]			&	0.39$^{+0.19}_{-0.13}$	&	0.39$^{+0.17}_{-0.12}$		&	0.37$^{+0.12}_{-0.11}$	& 0.43$^{+0.26}_{-0.17}$		\\
			k$_{gau}$[10$^{-3}$]	&	1.64$^{+0.87}_{-0.54}$		&	2.01$^{+0.93}_{-0.61}$			&	2.38$^{+1.05}_{-0.76}$		&	2.81$^{+2.01}_{-1.15}$				\\
			EW[eV]		& 32.4											& 36.8												& 43.2											& 53.2													\\
		\hline                                                                                                                                
		$\chi^2$					&	1528											& 1503												& 1421								    & 1401												\\
			 D.o.f					& 1512											& 1512												& 1512 								    &	1485												\\
		red. $\chi^2$			& 1.011											&	0.994												&	0.94											&	0.941												\\
		\hline\hline	
	\end{tabular}
\end{table} 

\section{Discussion}

\begin{figure}
	\centering
\includegraphics[width=\columnwidth]{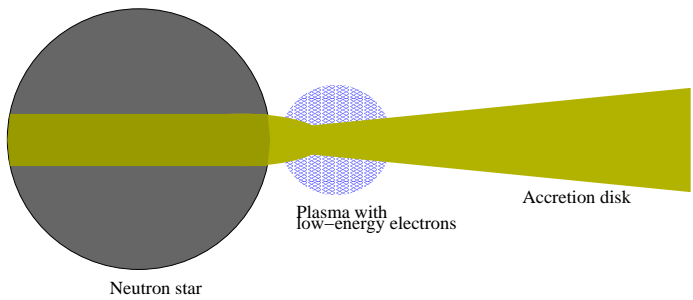} 
\caption{Possible geometry of the system. Illustration is based on \citet{Matsuoka12}. A Comptonization region of electrons with an energy of around 3\,keV is located close to the neutron star (marked in blue). The accretion takes place at the equatorial regions of the neutron star. The disk emits a multicolor blackbody. } 
	\label{Matsuoka}
\end{figure}

We have presented the analysis of four broad-band (0.2--24\,keV) \textit{BeppoSAX} observations of the LMXB 4U 1735-44. 
The broad-band spectra were modeled according to the eastern model based on \cite{Mitsuda84a, Mitsuda89a}. 
The continuum fit suggested the presence of a multicolor disk blackbody, originating from the disk, and a Comptonized component, most likely due to 
a corona of hot electrons close to the neutron star in the central region of the accretion disk. 
From the \textit{BeppoSAX} data, we also concluded that a reflection component can model the residuals of the continuum fit and especially the line-like features well. 
The most prominent feature was found at 6.7\,keV and originates from the iron K$_\alpha$ fluorescence line. 
Although the line is broad, relativistic blurring is not needed to model the line. This indicates that the production site of the observed line is not in the proximity of the neutron star. 
The inner radius of the accretion can be estimated from the normalization of the \textsc{diskbb} continuum model. 
We assumed an inclination angle of 60$^\circ$, which was found as an upper limit by \citet{Casares} and is also supported by the absence of dips in the light curve.
The hardening factor $\kappa$ \citep{Shimura95a} was set to 1.7 while the correction for inner boundary conditions was $\xi$=0.412 \citep{Kubota98a}.
Eventually, an inner disk radius of 18 to 23\,km was estimated from the \textit{BeppoSAX} observations.  
The inner disk radius is, therefore, between four to six times the Schwarzschild radius for a typical 1.4\,M$_{\odot}$ neutron star.
If we assume a neutron star radius between 10 and 15\,km \citep[based on the equation of state summarized in][]{Lattimer2007}, this indicates that the accretion disk neither reaches the surface of the compact object nor the innermost stable circular orbit (ISCO), which is three times the Schwarzschild radius.
However, if the mass of the neutron star is 2.0\,M$_{\odot}$, the Schwarzschild radius will be 5.9\,km and the 18\,km will match the ISCO. \\
\indent In our analysis of \textit{XMM-Newton} data, we analyzed the line profile, carefully taking into account pile-up and investigating the resulting effects. 
\cite{DiazTrigo10} claimed that pile-up can produce an asymmetric profile of the iron line.
In our analysis, we found no evidence for an asymmetry or broadening of the iron line caused by pile up.
On the contrary, we calculated a maximum line width of 0.430\,eV (equivalent width of 53.2\,eV) after removing the five innermost columns, compared with 0.389\,eV (32.4\,eV) with all columns.
Interestingly, the continuum spectrum became softer when more columns were excluded, which is expected because pile-up hardens the spectrum. 
The result obtained for the \textsc{diskline} model, a production site radius of more than 15 Schwarzschild radii ($>$60\,km), confirms that the line production does not take place in the innermost regions of the accretion disk. Moreover the \textsc{diskline} model produced an inclination value close to 60$^{\circ}$, which supports the assumption used for the \textit{BeppoSAX} data. \\
\indent According to our results, the model suggested by \citet{Sakurai12} and \citet[][see Figure \ref{Matsuoka}]{Matsuoka12} might well explain the emission geometry of 4U 1735-44.
In the soft state of the source, we observed a Comptonization region of electrons with an average energy of 3\,keV located close to the region between the neutron star and the disk. 
The seed photons are provided by the blackbody, which is emitted at the equatorial region of the neutron star and the inner parts of the accretion disk.
Furthermore, we observed a multicolor blackbody spectrum from the disk and the reflection component, which originates from regions of the disk far from the neutron star.
The direct emission from the neutron star could not be seen due to the Comptonization region marked blue in Figure \ref{Matsuoka}.
Although the model is based on the eastern model, the hot plasma close to the neutron star is similar to the accretion disk corona (ADC) proposed by \cite{Church04a}. 
However, the hot plasma region does not spread as far as the ADC.
It remains to be investigated why no line can be found in the analysis of Chandra data. 
The luminosity was similar to the observations presented in this work and the width of the line exceeded the upper limit given by \citet[][40\,eV]{Cackett09}.

\begin{acknowledgements}
	This work is supported by the Bundesministerium f\"ur Wirtschaft und Technologie through the Deutsches Zentrum für Luft und Raumfahrt (DLR, grants FKZ 50 OG 1001 and 50 OG 1301). We wish to thank Mrs Sarah Suchy for the careful reading of the manuscript.
\end{acknowledgements}

\end{document}